\documentclass[runningheads,a4paper]{cmmr2023}
\usepackage{times}
\usepackage{amssymb}
\usepackage{graphicx}
\usepackage{color}
\usepackage{float}
\usepackage{multirow}
\usepackage{hyperref}
\usepackage{subcaption}
\usepackage[table]{xcolor}
\newcommand{\keywords}[1]{\par\addvspace\baselineskip
\noindent\keywordname\enspace\ignorespaces#1}

\begin{document}

\mainmatter  % start of an individual contribution

\title{ProgGP: From GuitarPro Tablature Neural Generation To Progressive Metal Production}

\titlerunning{ProgGP: From Tablature Neural Generation To Progressive Metal Production}

\author{Jackson Loth\inst{1}, Pedro Sarmento\inst{1}, CJ Carr\inst{2},  Zack Zukowski\inst{2}
 \and Mathieu Barthet\inst{1} \thanks{This work is supported by the EPSRC UKRI Centre for Doctoral Training in Artificial Intelligence and Music (Grant no. EP/S022694/1). First and second author have equal contributions.}}
%
% if the names of the authors are too long for the running head, please use the format: AuthorA et al.
\authorrunning{Jackson Loth, Pedro Sarmento, CJ Carr, Zack Zukowski and Mathieu Barthet}

\institute{Queen Mary University of London, United Kingdom \and Dadabots, \url{https://dadabots.com/}\\ \email{j.j.loth@qmul.ac.uk}}


\maketitle

\begin{abstract}
%70-150 words
Recent work in the field of symbolic music generation has shown value in using a tokenization based on the GuitarPro format, a symbolic representation supporting guitar expressive attributes, as an input and output representation. We extend this work by fine-tuning a pre-trained Transformer model on ProgGP, a custom dataset of 173 progressive metal songs, for the purposes of creating compositions from that genre through a human-AI partnership. Our model is able to generate multiple guitar, bass guitar, drums, piano and orchestral parts. We examine the validity of the generated music using a mixed methods approach by combining quantitative analyses following a computational musicology paradigm and qualitative analyses following a practice-based research paradigm. Finally, we demonstrate the value of the model by using it as a tool to create a progressive metal song, fully produced and mixed by a human metal producer based on AI-generated music.

\end{abstract}

% Sections
\section{Introduction}

% \textcolor{red}{Say something as an introduction to the field of symbolic music generation (DONE) and practice based research. Also describe the genre of progressive metal.}
With advancements in computing power, new approaches to music generation have emerged. In recent years, deep learning has become a popular approach for automatic music generation, with research focusing on both the audio domain and the symbolic domain. This work extends previous work by Sarmento et al. \cite{sarmento2023gtr} using a symbolic music generation model trained on DadaGP, a symbolic music dataset consisting  26k songs of various genres \cite{Sarmento2021}. We follow here a practice-based research approach where a human expert music producer and music AI researchers collaborate to produce music based on machine-generated ouputs. We fine tuned the DadaGP-based model with a custom dataset of 173 progressive metal songs, which we refer to in this paper as ProgGP, with the intent of using the model to generate songs, which can be recorded and turned into a fully produced progressive metal song. The model used in this work generates music in the GuitarPro format, rather than formats such as MIDI, MusicXML and ABC seen in other symbolic music generation works \cite{Dong2020}. For guitar parts, GuitarPro not only encodes the pitch of each note, but also the location on a guitar fretboard where the note is meant to be played, as well as various expressive techniques (e.g. \textit{vibrato} and \textit{string bending}). We suggest that for certain musical genres, this format is very advantageous for a practice-based approach, as it provides much more information to an artist on how to perform the music that is generated, while still leaving room for creative interpretation. This paper presents the work that went into creating a brand new progressive metal song using neurally generated riffs and ideas that are relevant to the progressive metal genre. As per its main contributions, we highlight: (1) ProgGP, a manually curated progressive metal GuitarPro dataset made available to the community for research purposes; (2) a fine-tuned guitar tablature generative model for the creation of progressive metal tablatures; (3) heuristics for assessing whether generated music holds traits of the desired genre; (4) a practice-based research approach relying on a human-AI partnership where neurally-generated music is selected, edited, and integrated into a composition by a human producer. We also critically examine how to use neurally-generated music to foster creativity, inspire new ideas and improve the writing workflow of artists. We hope that this work will stir more research into human-AI interaction in the musical domain.
\section{Background}

\subsection{Symbolic Music Generation Using Deep Learning}

Recent advances in deep learning have led to promising results in the field of music generation \cite{sarmentoPerspectivesFutureSonic2021}, with techniques such as Variational Autoencoders (VAEs) \cite{Tan2020}, Generative Adversarial Networks (GANs) \cite{Dong2018}, Recurrent Neural Networks (RNNs) \cite{Meade2019} \cite{Sturm2016}, and Transformers \cite{AnnaHuang2019} being increasingly used. The Transformer model \cite{Vaswani2017} has enabled steep improvements in natural language processing (NLP) tasks and has been adapted for generating symbolic piano music in Huang et al.'s Music Transformer \cite{AnnaHuang2019}. Other notable works, such as Musenet \cite{christine_2019} and Pop Music Transformer \cite{Huang2020}, have further built on this approach to generate multi-instrument music and improve the generated music's rhythmic structure. However, the task of guitar tablature music generation has received limited research attention until the recent release of the DadaGP \cite{Sarmento2021} dataset, comprising songs in both GuitarPro format, a tablature edition software, and a dedicated textual token format. An initial example of guitar tablature generation work is Chen et al.'s fingerstyle guitar generator \cite{Chen2020}, despite not being based on the GuitarPro format. More recent works that explore the DadaGP dataset include GTR-CTRL \cite{sarmento2023gtr}, proposing a method for guitar tablature generation with control over instrumentation and musical genre, as well as LooperGP \cite{Adkins2023}, enabling to generate loopable music excerpts with applications for live coding performance.

\subsection{Practice-Based Research and Computer Music}

Many works deal with the notion of `practice' in research. Practice-based research is generally concerned with the knowledge gained through practice and the outcomes of that practice, while practice-led research leads to new understandings about practice itself \cite{candy2006practice}. Benford et al. describe this kind of research as consisting of three interconnected activities which inform each other in different ways: \textit{practice}, \textit{theory} and \textit{studies} \cite{benford2013performance}. However, they note challenges in conducting this research with balancing potentially different researcher and artist goals, as well as ethical concerns that can arise through artistic use of new technologies. Artistic uses of new technologies involving AI can be difficult due to the difficulty of prototyping new AI systems and the number of ways that AI can respond to users in different contexts \cite{yang2020re}. Amershi et al. \cite{amershi2019guidelines} provide guidelines on dealing with such unpredictable AI systems, mostly focusing on keeping the user informed on the system's capabilities and understanding its outputs. AI systems have seen use in musical practice-based research \cite{martelloni2021guitar} \cite{sturm2019machine} with the \textit{Folk-RNN} model by Sturm et al. being noted to have a number of impacts on musical creation such as a way to inspire ideas, break habits, and a sense of creating something that could not have been created otherwise.
\section{Practice-Based Research Methodology}

\subsection{Human-AI Partnership}
In this work, the first author, a music AI researcher and progressive metal producer, adopted the practice-based research approach described below:
\begin{enumerate}
    \item 
    Use a deep learning model to generate music in the style of the producer's preferred genre, progressive metal;

    \item 
    Evaluate the outputs of the model using a mixed method evaluation approach, combining objective metrics with subjective evaluation;

    \item 
    Craft a song using generated outputs based on outcomes from the evaluation;

    \item 
    Learn and record the song;

    \item 
    Analyse and reflect on the overall music production process.
\end{enumerate}

The work aims to better understand the successes and issues of the deep learning model in order to help the research community use and improve the model. We also publicly release the dataset used to fine-tune the deep learning model to support similar kinds of research. Finally, we develop a music production process which can be used to efficiently integrate neurally-generated content within a human composition. The artistic content that was recorded can be listened to online and could lead to public performances.

For the neural music generation, we use a model pre-trained on the DadaGP \cite{Sarmento2021} dataset, a dataset consisting of over 26k songs of various genres. The model is trained to produce songs in a tokenized symbolic format, which can be converted to the more commonly used GuitarPro format. This model is further fine-tuned on ProgGP, a curated dataset of progressive metal songs. This fine-tuned model can then be used to generate new songs in the style of progressive metal.
For clarification, we do not assess timbre quality aspects of progressive metal since we are working in the symbolic domain, despite timbre playing an important role in the genre (e.g. heavily distorted guitars, loud and punchy snare and kick drums, etc). However, we do take into account timbre identity through a distinction between distorted and clean guitars in our model.

\subsection{Fine-Tuning Dataset}
\begin{figure}
    \centering
    \includegraphics[scale=0.5]{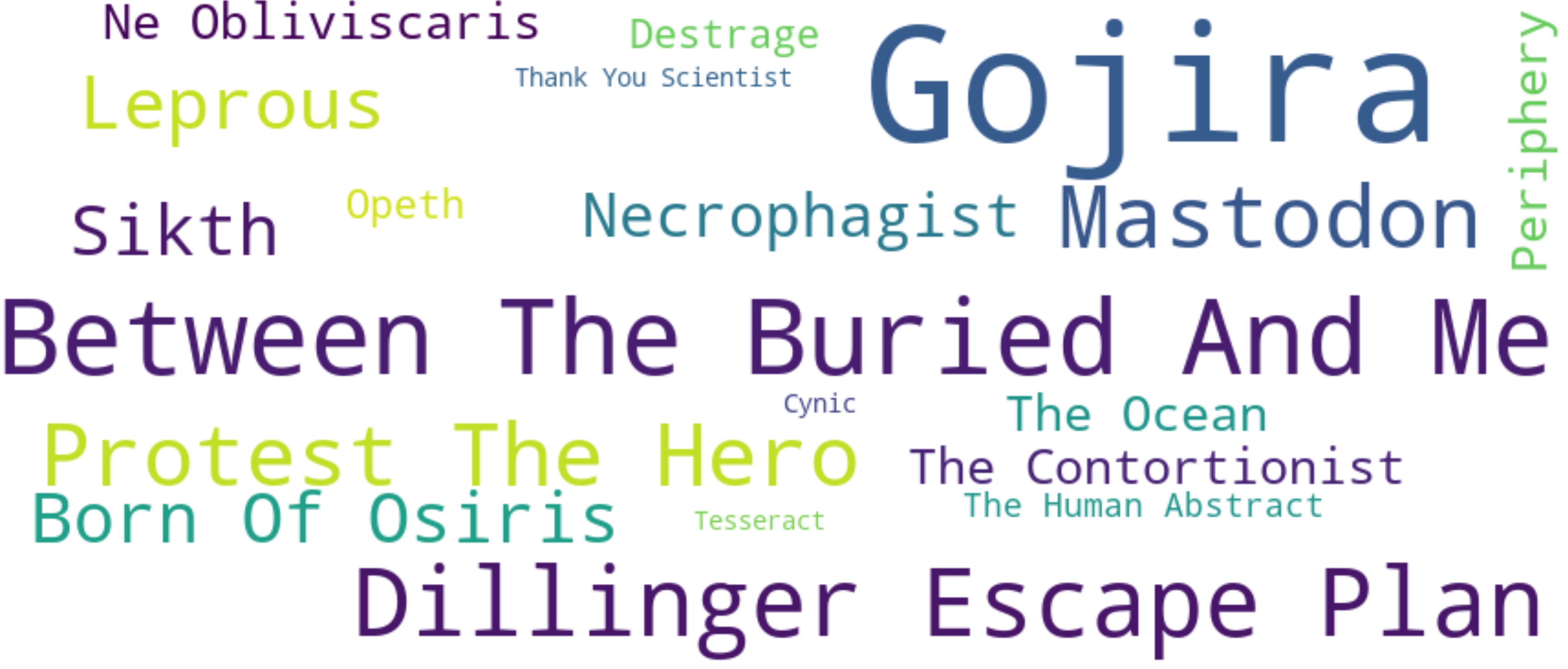}
    \caption{Word cloud representation of ProgGP's songs per artist distribution.}
    \label{fig:wordcloud}
\end{figure}

ProgGP, the fine-tuning dataset used in our experiments, consists of 173 songs largely from the progressive metal genre\footnote{Some songs included in the dataset are from adjacent genres (e.g. technical death metal). }. The songs were obtained using Songsterr\footnote{\url{https://www.songsterr.com/}}, a website that hosts GuitarPro files and allows playback using an web-based GuitarPro player. The tablatures (tabs) obtained from this website were not official tabs created by the artists of the songs, but rather created and maintained by the online community. Due to this, there is no guarantee that the tabs used in the dataset are perfectly accurate to the songs they are based on. However, each was verified to at least mostly capture the spirit of the original performance during the construction of the dataset. We limited the dataset to only songs in which the bass guitar and drums have also been transcribed, since the pre-trained model was trained on fully transcribed songs. This however limited the scope of the dataset, as many songs were only available with guitar transcriptions, rather than the full band. Additionally, the model only supports a few common guitar tunings, and only 6 and 7 string guitars. Many bands in this genre use more unique guitar tunings and/or 8 string guitars, so some artists that might be important in the genre of progressive metal may have limited songs or be absent entirely from the dataset. All this led to some artists dominating the dataset more than others. A word cloud representation of the artists used in the ProgGP dataset can be seen in Figure \ref{fig:wordcloud}. We made ProgGP\footnote{\url{https://github.com/otnemrasordep/ProgGP}} available upon request, together with a list of songs per artist. 

% \begin{table}[]
%     \centering
%     \caption{Number of songs used by each artist in the fine-tuning dataset. \textcolor{red}{Is this table necessary? Maybe make it a graph. Could also make it a footnote...}}
%     \begin{tabular}{|c|c|}
%         \hline
%         \textbf{Artist} & \textbf{\# of songs} \\
%         \hline
%         Born Of Osiris & 11\\
%         \hline
%         Between The Buried And Me & 23\\
%         \hline
%         The Contortionist & 5\\
%         \hline
%         Cynic & 1\\
%         \hline
%         Destrage & 3\\
%         \hline
%         Dillinger Escape Plan & 19\\
%         \hline
%         Gojira & 27\\
%         \hline
%         Leprous & 12\\
%         \hline
%         Mastodon & 17\\
%         \hline
%         Necrophagist & 8\\
%         \hline
%         Ne Obliviscaris & 5\\
%         \hline
%         Opeth & 3\\
%         \hline
%         Periphery & 5\\
%         \hline
%         Protest The Hero & 16\\
%         \hline
%         Sikth & 11\\
%         \hline
%         Tesseract & 1\\
%         \hline
%         Thank You Scientist & 1\\
%         \hline
%         The Human Abstract & 2\\
%         \hline
%         The Ocean & 5\\
%         \hline
%     \end{tabular}
%     \label{tab:artist_count}
% \end{table}

\subsection{Model Fine-Tuning}
The pre-trained model is based on the Transformer-XL \cite{Dai} architecture, a modified version of the original Transformer \cite{Vaswani2017} that is more capable of learning longer-term dependency. The pre-trained model used in our experiments was trained for 200 epochs on the DadaGP \cite{Sarmento2021} dataset. We trained the model on the fine-tuning dataset for an additional 65 epochs, at which the loss dropped low enough to trigger early stopping. Checkpoints were saved at every five epochs or training, resulting in 13 models at various stages of fine tuning. 

\subsection{Neural Generation}

A new song can be generated by feeding the model a prompt (set of instructions) in the form of a tokenized GuitarPro file. This will be the starting point of the generation, and the model will attempt to continue the song after the prompt. The tempo (in BPM) used for the generated song is taken from the prompt and the number of tokens to be generated is used as a parameter during inference. In DadaGP token format, a token can be a single note, rest, or expressive technique. Prompts used in the generation experiments ranged from a single note, a few measures from songs in the training set, and a few measures of songs not in the training set. The number of generated songs and the model from which to generate the songs can also be specified. Empirical analysis of the generated songs have allowed us to identify common structural patterns in generated songs, which we refer to as `sections', typically consisting of a \textit{riff} that is repeated one or more times with slight variations. The songs will typically start by repeating the notes from the prompt, with minor changes. It will then generate two or three sections afterward, each somewhat changing the feel of the song. While progressive metal songs can contain a large number of different riffs, they tend to build on one another and use references to musical motifs found throughout the song and throughout other songs by the same artist. Between The Buried And Me, a band with a large presence in ProgGP, is particularly well known for this \cite{hannan2019hearing}. This is a difficult thing to capture within a model however, as while the different sections seem to fit together naturally, they do not necessarily reference one another. Together with this submission, we release all the generated compositions on the undertaken experiments, cherry-picking some examples \footnote{Available at: \url{https://drive.google.com/drive/folders/1xaejTcUrPncE4hoyONhSzgS0a5TRo6G_?usp=share_link}}.
\section{Analysing AI-Generated Music}

We used a mixed method approach to better understand the outputs of the fine-tuned models, their strengths and weaknesses, and to help the producer select a model for further music production use. This was done by analysing the generated music from each model objectively through the use of common symbolic music metrics, as well as listening through many generated examples and analysing them subjectively in the context of the author's own knowledge of progressive metal.

\subsection{Objective Metrics}
\label{objectivemetrics}

Given the difficulties in assessing the quality of neurally-generated music without using a listening test, specially in the symbolic domain, we resorted on commonly used metrics from the literature, implemented in the MusPy package \cite{Dong2020}. For this evaluation, 173 songs were generated from each of the thirteen fine-tuned models, the same number of songs present within ProgGP, in order to maintain consistency when comparing the songs generated to the songs present in ProgGP. The prompt used in this analysis was a single low E note on guitar and bass guitar, and a kick and cymbal hit on drums. This was chosen in order to minimize the influence of the prompt as much as possible, as per the findings in \cite{sarmento2023gtr}. 

\begin{figure}[H]
    \centering
    \includegraphics[width=0.82\textwidth]{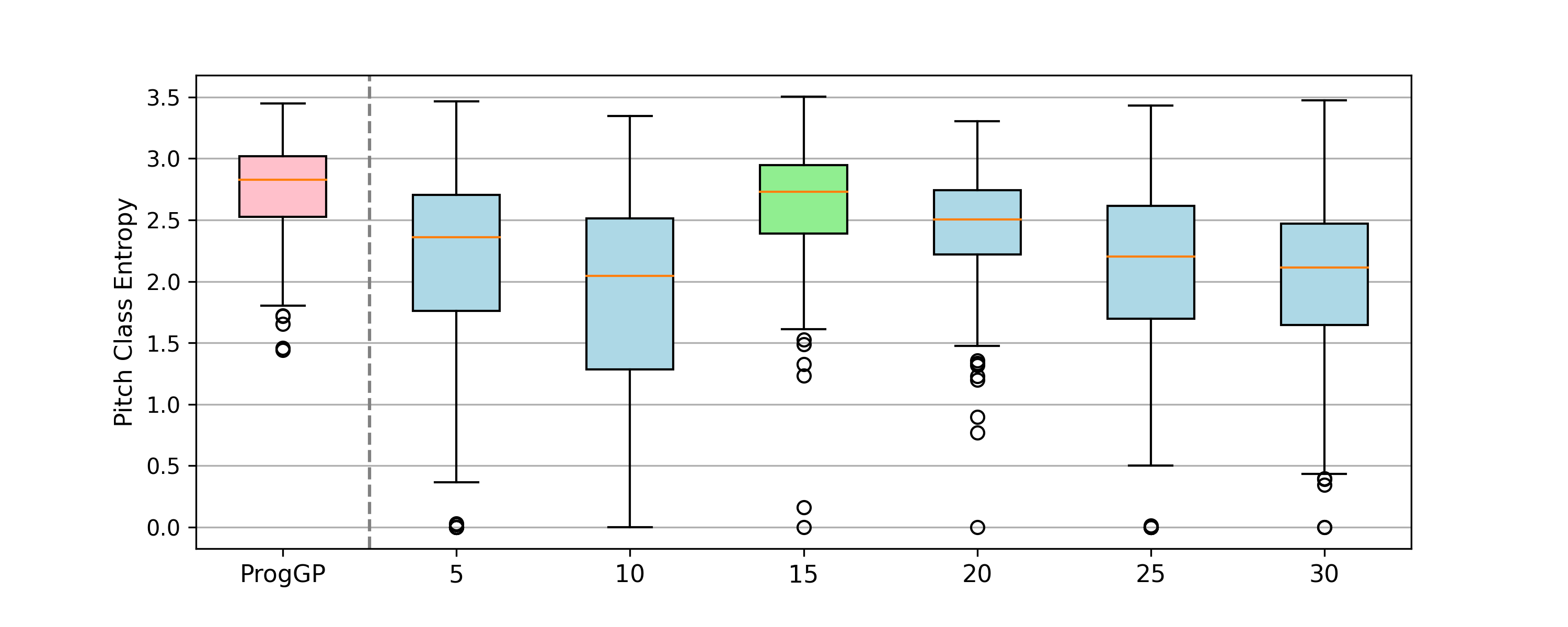}
    \caption{Pitch class entropy calculated for the songs in ProgGP (pink) and the generated songs from the fine-tuned models for different epochs (blue and green). Model with lowest KL-divergence highlighted (in green).}
    \label{fig:PCE}
\end{figure}

\begin{figure}[H]
    \centering
    \includegraphics[scale=0.27]{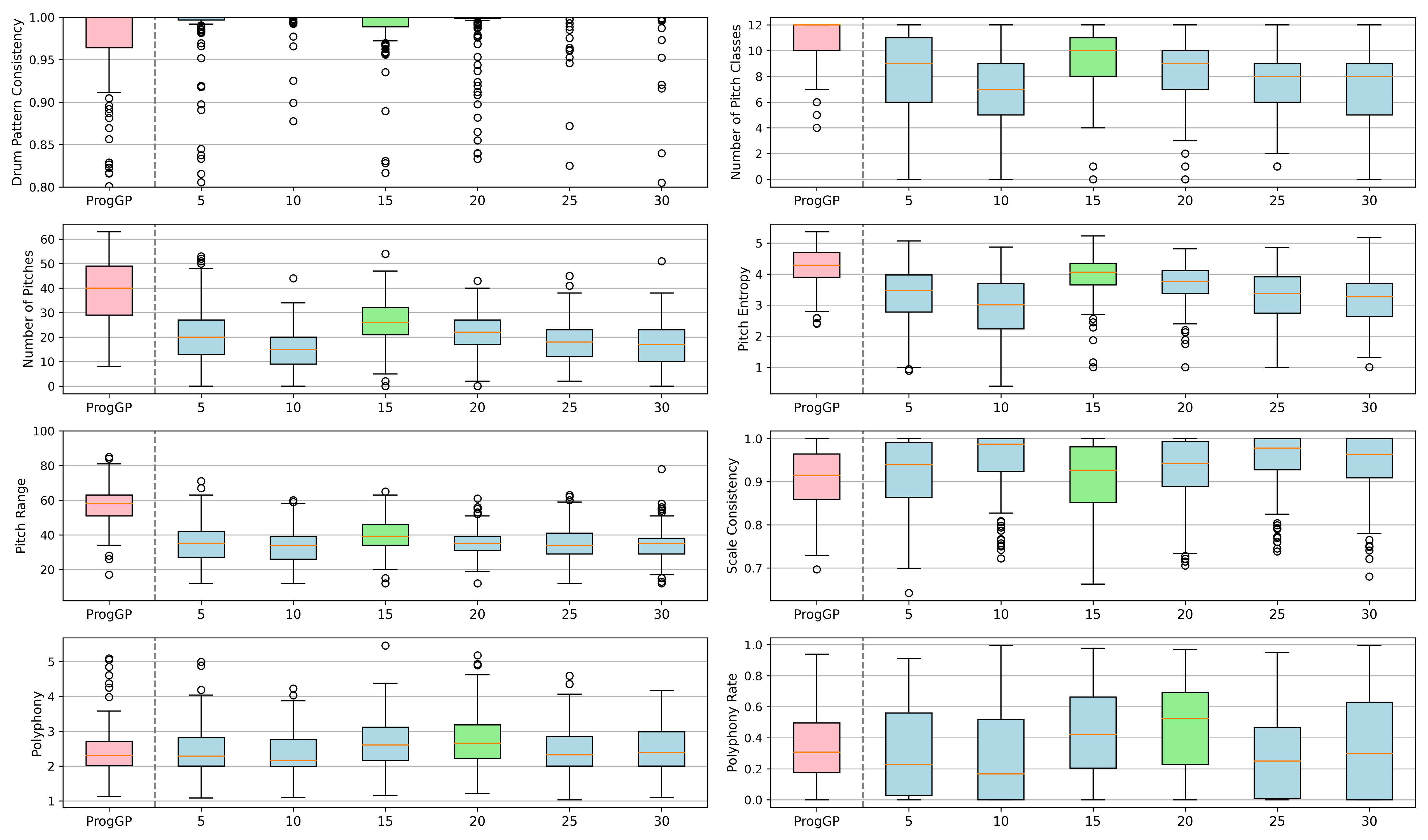}
    \caption{Metrics calculated for the songs in ProgGP (pink) and the generated songs for each fine-tuned model (blue and green). Model with lowest KLD highlighted (green).}
    \label{fig:metrics}
\end{figure}

In previous work, Sarmento et al. \cite{sarmento2023gtr} used \textbf{pitch class entropy} (PCE), a measure of the entropy of pitch classes used within a song, to evaluate their model. The PCE of the fine tuned models can be seen in Figure \ref{fig:PCE} (to ease visualization, we omit plots from models after epoch 30). The models fine-tuned for 15 and 20 epochs seem to have a distribution closer to ProgGP. The models fine-tuned for 5 and 10 epochs and beyond 20 epochs generally have a lower mean than the 15 and 20 epoch models. We hypothesize that this could be due to overfitting, causing the model to get stuck on certain sections or notes and repeating them, something seen in the generated songs by the more fine-tuned models. This would lower the pitch class entropy of a model's outputs rather than push it closer to that of the training data which is higher. The rest of the metrics can be seen in Figure \ref{fig:metrics}. They include \textbf{drum pattern consistency} (DPC), \textbf{number of pitch classes} (NPC), \textbf{number of pitches} (NP), \textbf{pitch entropy} (PE), \textbf{pitch range} (PR), \textbf{scale consistency} (SC), \textbf{polyphony} (Pol) and \textbf{polyphony rate} (PolR). These metrics, while not necessarily giving a definitive idea of the performance of a model, help us understand how the output of certain models matches the training data. They also give an idea of certain characteristics of the music that each model tends to generate. An in-depth definition of each can be found in MusPy's package documentation\footnote{\url{https://salu133445.github.io/muspy/metrics.html}}. 

The Kullback-Leibler divergence (KLD), a measure of relative entropy between the true probability distribution and a sample probability distribution, was calculated for each of the fine-tuned models (ProgGP is used as groundtruth to compared against generated songs). The KLD results can be seen in Table \ref{tab:KLDiv}.

\begin{table}[H]
    \centering
    \setlength{\arrayrulewidth}{1pt}
% \arrayrulecolor{black}
\setlength{\tabcolsep}{5pt}
\renewcommand{\arraystretch}{1.0}
    \caption{KLD scores for each fine-tuned model against ProgGP. Bold and green coloring indicates the lowest KLD per column.}
    \vspace{0.05cm}
    \scalebox{0.82}{
    \begin{tabular}{|c||c|c|c|c|c|c|c|c|c|}
    \hline
          \textbf{Epoch} & \textbf{PCE} & \textbf{DPC} & \textbf{NPC} & \textbf{NP} & \textbf{PE} & \textbf{PR} & \textbf{Pol} & \textbf{PolR} & \textbf{SC} \\
\hline
\hline
5 & 0.473 & 0.513 & 0.608 & 0.799 & 0.638 & 0.762 & 0.497 & 0.495 & 0.263 \\
\hline
10 & 0.665 & 0.696 & 1.599 & 1.052 & 0.800 & 0.845 & 0.570 & 0.573 & 0.433 \\
\hline
15 & \cellcolor{green!25}\textbf{0.262} & \cellcolor{green!25}\textbf{0.442} & \cellcolor{green!25}\textbf{0.491} & \cellcolor{green!25}\textbf{0.746} & \cellcolor{green!25}\textbf{0.442} & \cellcolor{green!25}\textbf{0.591} & 0.365 & 0.353 & \cellcolor{green!25}\textbf{0.216} \\
\hline
20 & 0.425 & 0.478 & 0.999 & 0.914 & 0.616 & 1.062 & \cellcolor{green!25}\textbf{0.301} & \cellcolor{green!25}\textbf{0.247} & 0.286 \\
\hline
25 & 0.673 & 0.596 & 1.641 & 0.998 & 0.670 & 0.912 & 0.484 & 0.559 & 0.491 \\
\hline
30 & 0.707 & 0.640 & 1.200 & 1.043 & 0.851 & 1.054 & 0.400 & 0.509 & 0.312 \\
\hline
35 & 0.625 & 0.625 & 1.144 & 0.939 & 0.743 & 0.974 & 0.376 & 0.493 & 0.376 \\
\hline
40 & 0.480 & 0.611 & 1.050 & 0.970 & 0.717 & 1.121 & 0.513 & 0.544 & 0.274 \\
\hline
45 & 0.702 & 0.746 & 1.554 & 1.059 & 0.910 & 1.089 & 0.420 & 0.486 & 0.336 \\
\hline
55& 0.648 & 0.679 & 1.510 & 1.040 & 0.813 & 1.092 & 0.517 & 0.504 & 0.317 \\
\hline
55 & 0.595 & 0.690 & 1.358 & 1.039 & 0.818 & 1.092 & 0.471 & 0.485 & 0.346 \\
\hline
60 & 0.681 & 0.677 & 1.513 & 1.018 & 0.816 & 1.157 & 0.579 & 0.575 & 0.375 \\
\hline
65 & 0.757 & 0.730 & 2.069 & 1.126 & 0.842 & 1.041 & 0.394 & 0.484 & 0.379 \\
\hline
    \end{tabular}}
    \label{tab:KLDiv}
\end{table}
%Most of the models have very high scale consistency, despite ProgGP having much more variance in this area. Scale consistency measures the highest rate at which notes are within either major or natural minor scales, for any possible root note. As progressive metal tends to have a focus on experimentation, it makes sense that some songs might have low scale consistency. However, the fine-tuned models seem to have trouble matching that aspect of the songs in ProgGP. We hypothesize that this might be due to effects of the prompt used in this analysis.
% GuitarPro also supports expressive playing techniques such as string bends, palm mutes, vibrato and more. These are present within ProgGP and the songs generated by the models. Figure \ref{} shows the distribution of these within ProgGP and the generated songs from each of the fine-tuned models. \textcolor{red}{Need to do these calculations and talk about them}.

The model fine-tuned for 15 epochs scores the lowest for most metrics. The only exceptions are polyphony and polyphony rate, in which the model fine-tuned for 20 epochs scores the lowest. This is expected given that the model trained for 15 epochs seems to be more similar to ProgGP for most of the metrics than the other models.

\subsection{Subjective Analysis}

Subjectively evaluating generated progressive metal songs first requires a definition of progressive metal. This definition is hard to specify, as music genres are not always straightforward. Nevertheless, there are a number of tropes that progressive metal songs tend to have. Robinson \cite{robinson2019exploration} describes several of these such as polyrhythms, syncopated chugging on low notes and uncommon time signatures. These can be seen in many generated songs, particularly uncommon time signatures and syncopated rhythms. Similarly to the conclusions from GTR-CTRL \cite{sarmento2023gtr}, we empirically found that the prompt has a reasonably large amount of influence over the generated song, but this varies between songs. The model tends to only generate notes for instruments contained in the prompt (e.g if there exists two guitars, one bass guitar and drums within the prompt, the model will only generate new notes for those instruments). It does however occasionally generate an extra guitar or keyboard track (\href{https://drive.google.com/file/d/1x-9MJg5UK5zWNm5CJns730T0tlBBiTXv/view?usp=share_link}{\underline{id-00}})\footnote{Song ids are hyperlinked to facilitate listening.}, but these scenarios were found to be rare. Generated guitar parts for multiple guitar tracks tend to be mostly identical, mirroring the recording technique of two guitars playing identical parts in order to create width in a song mix. Interestingly however, the model will sometimes generate a harmony for a particular guitar line where one guitar plays some kind of melodic line and the other playing the same line with the pitch shifted (\href{https://drive.google.com/file/d/1e0c9X-X8im9LKlbicVTOMygCj-1Lg-4K/view?usp=share_link}{\underline{id-01}}). It also occasionally generates guitar solos and rhythmic accompaniment (\href{https://drive.google.com/file/d/18m38SUPTeHysIwgGuiVxO3WhK6iYGT4C/view?usp=share_link}{\underline{id-002}}), with one guitar playing low-pitched chords while the other plays fast single high-pitched notes. The model generates very impressive drum parts in addition to the guitar and bass guitar (\href{https://drive.google.com/file/d/1xJGQZNHNaGU18uBEQ_9O01yV1oq7BPHr/view?usp=share_link}{\underline{id-03}}). The timing of the kick drum consistently lines up with the notes of the bass guitar (\href{https://drive.google.com/file/d/1sFeJTB5Gei9GYnTmeLoRrKmRi1TEjgZR/view?usp=share_link}{\underline{id-04}}). Additionally, several common drum beats heard in many metal songs can be generated (e.g. blast beats (\href{https://drive.google.com/file/d/1ScBriV67HH-KTdrYebSjyWlnl7GUDSt1/view?usp=share_link}{\underline{(id-05}})). Many songs also feature drum fills at the end of a section before transitioning into a new section. It is possible that the model excels at generating drum parts due to the limited number of possible notes compared to pitch-based instruments such as guitar and bass guitar. This being said, the generated drum parts would likely need further editing if used in an actual song in order to convey more of the nuance heard in progressive metal drumming. 
% They are found to be better at conveying ideas rather than being actual final drum parts.

\section{Song Production}

A short progressive metal song was recorded, produced and mixed using one of the fine-tuned models to generate the initial musical ideas and song structure. This was done by the first author, himself a progressive metal producer and music AI researcher. The intention with this production was to utilize the generated songs as a way to bolster creativity and inspire ideas for music in a way in which the artist's creativity can still be applied to integrate the generated content into a song of their own. Section \ref{process} describes a high level overview of the song creation process using the AI system in collaboration with a music producer, while Section \ref{songanalysis} presents a detailed analysis of the generated song and what was changed in order to suit the production.

\subsection{Process}
\label{process}

The process of creating the song can be broken into the following steps:
\begin{enumerate}
    \item 
    A prompt is selected and songs are generated using one a fine-tuned model. One is chosen to be the starting point of the song based on how it inspires the producer.

    \item 
    The generated song is loaded into a guitar tab reader software (e.g. GuitarPro).

    \item 
    Drums and bass are exported to MIDI format and loaded into a digital audio workstation (DAW), along with appropriate virtual instruments.

    \item 
    The guitar parts are learned by the guitarist producer from the generated guitar tab and subsequently recording in the DAW. During the recording of the guitar, changes can be made to suit the producer's idea of the direction of the song.

    \item 
    The drum and bass guitar MIDI are edited to suit any changes made to the guitar, or to better serve the song. This may be done in conjunction with the previous step and may require some back and forth in order to fully develop the song.
\end{enumerate}

These steps can be repeated as many times as desired to build out a complete song. They may even be skipped if the producer is inspired by the ideas to create their own parts based on what was already generated. Virtual instruments for the bass guitar and drums are not strictly needed, but can assist in speeding up the workflow. It was found that this strategy allowed for a song to be developed quickly and minimized any extra work that may distract from creativity (e.g. having to record bass guitar parts in addition to the guitar parts or manually programming drum parts). In the next section we focus on a particular example generated using the first two measures of ``Stabwound" by Necrophagist as the prompt. The song was generated using the model fine-tuned on ProgGP for 15 epochs. The structure of the generated song was not changed, as we felt that it had many interesting qualities. The guitar, drums and bass were changed slightly to better fit the vision that the generated song inspired. Additional sounds such as synths, organs and impact samples were also added to flesh out the song and increase interest in the production. The final mix and the original generated song in both PDF and GuitarPro format are available online\footnote{Available at: \url{https://drive.google.com/drive/folders/1y2xX3WIQeOz6Z8FoN2VP3kzWvOqYk8QI?usp=sharing}}.

\subsection{Song and Production Analysis}
\label{songanalysis}

 The first section of the song is made up of an idea which takes up 4 measures. This idea is repeated with the second repetition skipping the first measure of the motif and adding on a new lick in the final measure which helps transition the section into the next one. Each repetition has a similar structure: three measures of 4/4 and a final measure with an odd time signature. The first repetition adds a 5/4 time signature to the end, while the second section uses a 6/4 time signature. Time signature changes are common in progressive metal \cite{robinson2019exploration}, and it is interesting to see the model generate this time signature change in both repetitions of the initial idea without simply repeating the idea. The changes in the second repetition of the idea feel like something a real songwriter might intentionally write, as if the model is building on the initial idea to create more excitement before the next section. The second section shows off a major flaw of the model: it does not always generate tabs or ideas that can be reasonably played by a human. Since a specific pitch can be played at multiple different areas of the guitar fretboard, tabs specify exactly which fret and string a note should be played on. However, the model will sometimes generate fretboard locations that are very unnatural to play by a guitarist. The tabs had to be slightly modified in order to record this section, however keeping the same notes. The main idea in this section is a repeated line of seven 8th notes followed by a chromatic note run and a lick that changes the modality from major to minor halfway through. It is difficult to know if this is something the model learned through training or if this note selection was more random. The section ends with four simple chords to transition into the next one. These were changed to be more dissonant chords in the recorded version. The final section is another repeated riff of seven notes used in a slightly more musical way than the previous section. Each repetition uses the same relative intervals between notes to outline two different chords, F\# minor and G\# minor. It then ends the section with two measures of 4/4, helping the song end in a slightly more familiar and natural way. A lick from the previous section is used in this ending in the tab, which helps tying the two sections together and increases cohesion.

\begin{figure}
    \centering
    \includegraphics[scale=0.55]{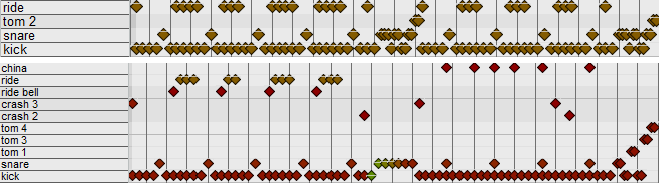}
    \caption{Original generated drum MIDI (top) vs. the final edited drum MIDI (bottom).}
    \label{fig:MIDI}
\end{figure}

While the structures and guitar riffs remained largely unchanged, the drums did not support the rest of the song as well as they could. While many generated songs have impressive sounding drums, the drum parts generated in this particular song did not quite hold up to professional standards. The first section  mostly had a snare fill which did not enhance the interesting aspects of the guitar and bass parts. This was changed to use a more steady snare hit and cymbals on the downbeats of the measure. A stack cymbal was used in the first repetition, but was changed to a china cymbal in the second repetition to add excitement to the changes between the two repetitions. A drum fill was also added in during the last few beats of the section to help highlight the transition between the two sections. The drums for the second section were mostly the same as the generated drums. The generated snare drum placement in this section accents the 7/4 time signature. However, the ride cymbals in the second repetition were changed to china cymbals which hit on the downbeats of the measure, and the kick drum was changed to be constant eight notes. This was done to push the energy up as the section finishes. The drums in the final section were kept mostly unchanged, with a small change to the drum fill at the end. A comparison from a section of the song of the originally generated MIDI and the edited MIDI can be seen in Figure \ref{fig:MIDI}. The process showed that while the model can excel at generating inspiring progressive metal ideas, a decent amount of work is still needed to make the ideas playable and professional sounding. Drums in particular, while containing good initial ideas, need a lot of editing to make them sound natural and support the ideas in the guitar and bass guitar parts. It is not as simple as directly importing the drum and bass MIDI from the generated song, a human producer is still required to make the ideas into something that is satisfying to listen to and convey emotion properly. That being said, the entire writing and production process only took three to four hours over two sessions, with most of the time being spent practicing the guitar parts in order to play them to a sufficient level for recording. The producer felt that the AI system helps inspiring new ideas and producing a good sounding demo extremely quickly, with an amazing level of detail in both the kinds of notes generated and song structure. It is easy to imagine combining multiple generated ideas together in this way to produce a full length song.
\section{Conclusion and Future Work}

We have presented a deep learning model capable of generating songs in the style of progressive metal. We released ProgGP, a symbolic music dataset consisting of 173 progressive metal songs, which was constructed and used to fine-tune a pretrained transformer model. The models fine-tuned for only a relatively small number of epochs, such as 15 and 20 epochs, produce interesting results and are shown to exemplify traits of the fine-tuning data in nine different symbolic music metrics. This analysis was used to inform the selection of a generated song, which was then turned into a full progressive metal production. Finally, we presented an analysis of the generated song and how it was used to augment the producer's own creativity. We hope to continue this collaboration between human musicians and the AI system in a possible professionally recorded album and live performance of AI-assisted progressive metal songs.

\bibliographystyle{splncs04}
\bibliography{mybibliography}

\end{document}